\documentclass[11pt]{article}
\usepackage{graphicx}
\usepackage{amsmath}
\usepackage{amssymb}
\usepackage{caption2}
\begin{document}
\bibliographystyle{prsty}
\begin{center}
{\large {\bf \sc{  Another tetraquark  structure  in the $\pi^+ \chi_{c1}$ invariant mass distribution }}} \\[2mm]
Zhi-Gang Wang \footnote{E-mail,wangzgyiti@yahoo.com.cn.  }     \\
 Department of Physics, North China Electric Power University,
Baoding 071003, P. R. China
\end{center}

\begin{abstract}
In this article, we assume that there exists a scalar hidden charm
tetraquark state in the $\pi^+ \chi_{c1}$ invariant mass
distribution, and study its mass using the QCD sum rules. The
numerical result
  $M_{Z}=(4.36\pm0.18)\,\rm{GeV}$
 is  consistent with the mass of the
$Z(4250)$. The  $Z(4250)$ may be a tetraquark state, other
possibilities, such as a hadro-charmonium resonance and a
$D_1^+\bar{D}^0+ D^+\bar{D}_1^0$ molecular state  are not excluded.
\end{abstract}



\section{Introduction}

Recently the Belle collaboration  reported the first observation of
two resonance-like structures  in the $\pi^+\chi_{c1}$ invariant
mass distribution near $4.1 \,\rm{GeV}$ in the exclusive
$\bar{B}^0\to K^- \pi^+ \chi_{c1}$ decays \cite{Belle-chipi}.  The
Breit-Wigner
 masses and the widths are about
$M_1=4051\pm14^{+20}_{-41} \,\rm{MeV}$,
$\Gamma_1=82^{+21}_{-17}$$^{+47}_{-22}\, \rm{MeV}$,
$M_2=4248^{+44}_{-29}$$^{+180}_{-35}\,\rm{MeV}$ and
$\Gamma_2=177^{+54}_{-39}$$^{+316}_{-61}\,\rm{MeV}$ (thereafter we
will denote them as $Z(4050)$ and $Z(4250)$ respectively).
 The significance of each of the $\pi^+\chi_{c1}$ structures
exceeds $5 \sigma$, including the effects of systematics from
various fit models. Their quark contents must be some special
combinations of $c\bar{c} u\bar{d}$, just like the $Z(4430)$, they
cannot be the conventional mesons \cite{Belle-z4430}.

The $Z$ (denote the $Z(4050)$ and $Z(4250)$) may be  loosely
deuteron-like bound states  (molecules) of the charm mesons or
compact nucleon-like bound states of the diquark-antiquark pair.
 The spins of the $Z$  are not determined yet, they can be scalar or vector
states.

In the meson-exchange model, the $Z(4050)$ is probably a loosely
molecular state $D^*\bar{D^*}$ with $J^{P}=0^+$ \cite{Liu0808} and
the $Z(4250)$ is unlikely  an $S$-wave ${\rm D_1D}$ or ${\rm
D_0D^{*}}$ molecular state \cite{Ding2008}, while the $SU(3)$ chiral
quark model indicates that the $Z(4050)$ is unlikely  an $S$-wave
$D^*\bar{D}^*$ molecular state \cite{Zhang2008}. In
Ref.\cite{SHLee0808,SHLee08082}, the authors study
   the mesons  $Z(4050)$ and $Z(4250)$  as
the $D^{*}\bar{D}^{*}$ and $D_1^+\bar{D}^0+ D^+\bar{D}_1^0$
molecular states with  $J^P=0^+$ and $J^P=1^-$ respectively using
the QCD sum rules, and draw the conclusion that the
$D^{*}\bar{D}^{*}$ state is probably a virtual state which is not
related with the $Z(4050)$  and the $Z(4250)$ is a possible
$D_1^+\bar{D}^0+ D^+\bar{D}_1^0$ molecular state. In a relativistic
quark model, the $Z(4250)$ can be tentatively interpreted as the
charged  $P$-wave $1^-$  tetraquark state $S\bar{S}$ or as the
$P$-wave $0^-$ tetraquark state $(S\bar{A}\pm \bar{S}A)/\sqrt{2}$
 \cite{Ebert2008}, where the $S$ and $A$ denote the
scalar and axial vector diquarks respectively.

The colored objects (diquarks) in a  confining potential can result
in a copious spectrum, there maybe  exist  a series of orbital
angular momentum excitations; while the colorless objects (mesons)
bound by a short range potential (through meson-exchange) should
have a very limited spectrum. In the heavy quark limit, the $c$
quark can be viewed as a static well potential, and binds  the light
quark $q$ to form a diquark in the color antitriplet channel. We
take the diquarks as the basic constituents   following  Jaffe and
Wilczek \cite{Jaffe2003,Jaffe2004}.  The heavy tetraquark system
could be described  by a double-well potential with  two light
quarks $q'\bar{q}$ lying in the two wells respectively.

In
Refs.\cite{Maiani2004,Maiani20042,Maiani2005,Maiani2008,Polosa0902},
Maiani et al take the diquarks as the basic constituents, examine
the rich spectrum of the
 diquark-antidiquark states  from the constituent diquark masses and the spin-spin
 interactions, and try to  accommodate some of the newly observed charmonium-like resonances not
 fitting a pure $c\bar{c}$ assignment. The predictions depend heavily on  the assumption that the light
 scalar mesons $a_0(980)$ and $f_0(980)$ are tetraquark states,
 the  basic  parameters (constituent diquark masses) are
 estimated thereafter.

In Ref.\cite{Wang0807}, we assume  that the hidden charm mesons
$Z(4050)$ and $Z(4250)$ are vector tetraquark states,  and study
their masses using the QCD sum rules. The numerical results
indicate that the masses of the vector hidden charm tetraquark
states are about $M_{Z}=(5.12\pm0.15)\,\rm{GeV}$ or
$M_{Z}=(5.16\pm0.16)\,\rm{GeV}$, which are  inconsistent with the
experimental data and also much larger than the predictions of the
constituent diquark model
\cite{Maiani20042,Maiani2005,Maiani2008,Polosa0902}.

The diquarks have five Dirac tensor structures, scalar $C\gamma_5$,
pseudoscalar $C$, vector $C\gamma_\mu \gamma_5$, axial vector
$C\gamma_\mu $  and tensor $C\sigma_{\mu\nu}$. The structures
$C\gamma_\mu $ and $C\sigma_{\mu\nu}$ are symmetric, the structures
$C\gamma_5$, $C$ and $C\gamma_\mu \gamma_5$ are antisymmetric. The
attractive interactions of one-gluon exchange favor  formation of
the diquarks in  color antitriplet $\overline{3}_{ c}$, flavor
antitriplet $\overline{3}_{ f}$ and spin singlet $1_s$
\cite{GI1,GI2}. The scalar hidden charm tetraquark states may have
smaller masses than the corresponding vector states.

The mass is a fundamental parameter in describing a hadron, in order
to identify  the $Z(4050)$ and $Z(4250)$ as tetraquark states, we
must prove that the masses of the corresponding tetraquark states
lie in the region $(4.1-4.3)\, \rm{GeV}$. Furthermore, whether or
not there exist such hidden tetraquark configurations is of great
importance itself, because it provides a new opportunity for a
deeper understanding of the low energy QCD.

In this article, we assume that  there exists  a scalar hidden charm
tetraquark state in the $\pi^+ \chi_{c1}$ invariant mass
distribution, and construct the $C\gamma_5-C \gamma_5$ type current
$J_1(x)$ and $C-C $ type current $J_2(x)$ (and their superposition
$J(x)$) to interpolate it,
\begin{eqnarray}
J_1(x)&=& \epsilon^{ijk}\epsilon^{imn}u_j^T(x) C\gamma_5 c_k(x)
\bar{c}_m(x) \gamma_5  C \bar{d}_n^T(x)\, , \\
J_2(x)&=& \epsilon^{ijk}\epsilon^{imn}u_j^T(x) C c_k(x) \bar{c}_m(x)
  C \bar{d}_n^T(x)\, , \\
  J(x)&=&cos\theta J_1(x)+sin\theta J_2(x) \, ,
\end{eqnarray}
where the $i$, $j$, $\cdots$, $n$ are color indexes; then study its
mass using the QCD sum rules \cite{SVZ79,Reinders85}. The hidden
charm mesons $X(3872)$, $Y(4260)$, $Y(4350)$, $Y(4660)$, $Z(4430)$
have also been studied with the QCD sum rules as the tetraquark or
molecular states \cite{Narison07,Lee07-sum,SR4430,SR4260}.

 In the QCD sum rules, the operator product expansion is used to expand
the time-ordered currents into a series of quark and gluon
condensates which parameterize the long distance properties of  the
QCD vacuum. Based on the quark-hadron duality, we can obtain copious
information about the hadronic parameters at the phenomenological
side \cite{SVZ79,Reinders85}.

The article is arranged as follows:  we derive the QCD sum rules for
the mass of  the $Z$  in section 2; in section 3, numerical results
and discussions; section 4 is reserved for conclusion.

\section{QCD sum rules for  the tetraquark state $Z$ }
In the following, we write down  the two-point correlation function
$\Pi(p)$  in the QCD sum rules,
\begin{eqnarray}
\Pi(p)&=&i\int d^4x e^{ip \cdot x} \langle
0|T\left\{J(x)J^{\dagger}(0)\right\}|0\rangle \, ,
\end{eqnarray}
we choose  the scalar current $J(x)$ to interpolate the tetraquark
state $Z$.

We can insert  a complete set of intermediate hadronic states with
the same quantum numbers as the current operator $J(x)$ into the
correlation function $\Pi(p)$  to obtain the hadronic representation
\cite{SVZ79,Reinders85}. After isolating the ground state
contribution from the pole term of the $Z$, we get the following
result,
\begin{eqnarray}
\Pi(p)&=&\frac{\lambda_{Z}^2}{M_{Z}^2-p^2} +\cdots \, \, ,
\end{eqnarray}
where the pole residue (or coupling) $\lambda_Z$ is defined by
\begin{eqnarray}
\lambda_{Z} &=& \langle 0|J(0)|Z(p)\rangle \, .
\end{eqnarray}

 In the following, we briefly outline  the operator product
expansion for the correlation function $\Pi(p)$  in perturbative
QCD. The calculations are performed at  the large space-like
momentum region $p^2\ll 0$. We write down the "full" propagators
$S_{ij}(x)$ and $C_{ij}(x)$ of a massive quark in the presence of
the vacuum condensates firstly \cite{Reinders85},
\begin{eqnarray}
S_{ij}(x)&=& \frac{i\delta_{ij}\!\not\!{x}}{ 2\pi^2x^4}
-\frac{\delta_{ij}m_q}{4\pi^2x^2}-\frac{\delta_{ij}}{12}\langle
\bar{q}q\rangle +\frac{i\delta_{ij}}{48}m_q
\langle\bar{q}q\rangle\!\not\!{x}-\frac{\delta_{ij}x^2}{192}\langle \bar{q}g_s\sigma Gq\rangle\nonumber\\
&& +\frac{i\delta_{ij}x^2}{1152}m_q\langle \bar{q}g_s\sigma
 Gq\rangle \!\not\!{x}-\frac{i}{32\pi^2x^2} G^{ij}_{\mu\nu} (\!\not\!{x}
\sigma^{\mu\nu}+\sigma^{\mu\nu} \!\not\!{x})  +\cdots \, ,
\end{eqnarray}
\begin{eqnarray}
C_{ij}(x)&=&\frac{i}{(2\pi)^4}\int d^4k e^{-ik \cdot x} \left\{
\frac{\delta_{ij}}{\!\not\!{k}-m_c}
-\frac{g_sG^{\alpha\beta}_{ij}}{4}\frac{\sigma_{\alpha\beta}(\!\not\!{k}+m_c)+(\!\not\!{k}+m_c)\sigma_{\alpha\beta}}{(k^2-m_c^2)^2}\right.\nonumber\\
&&\left.+\frac{\pi^2}{3} \langle \frac{\alpha_sGG}{\pi}\rangle
\delta_{ij}m_c \frac{k^2+m_c\!\not\!{k}}{(k^2-m_c^2)^4}
+\cdots\right\} \, ,
\end{eqnarray}
where $\langle \bar{q}g_s\sigma Gq\rangle=\langle
\bar{q}g_s\sigma_{\alpha\beta} G^{\alpha\beta}q\rangle$  and
$\langle \frac{\alpha_sGG}{\pi}\rangle=\langle
\frac{\alpha_sG_{\alpha\beta}G^{\alpha\beta}}{\pi}\rangle$, then
contract the quark fields in the correlation function $\Pi(p)$ with
Wick theorem, and obtain the result:
\begin{eqnarray}
\Pi(p)&=&i\epsilon^{ijk}\epsilon^{imn}\epsilon^{i'j'k'}\epsilon^{i'm'n'}\int
d^4x e^{ip \cdot x} \left\{ cos^2\theta Tr\left[\gamma_5 C_{kk'}(x)
\gamma_5 C S_{jj'}^T(x) C
\right]\times \right.\nonumber\\
&&Tr\left[\gamma_5 C_{m'm}(-x) \gamma_5 C S_{n'n}^T(-x) C
 \right]+sin^2\theta Tr\left[ C_{kk'}(x)
 C S_{jj'}^T(x) C \right]\times\nonumber\\
 &&\left. Tr\left[  C_{m'm}(-x)   C S_{n'n}^T(-x) C
 \right] \right\}\, .
\end{eqnarray}

Substitute the full $u$, $d$ and $c$ quark propagators into the
correlation function  $\Pi(p)$ and complete  the integral in the
coordinate space, then integrate over the variables in the momentum
space, we can obtain the correlation function $\Pi(p)$ at the level
of the quark-gluon degrees  of freedom.

We carry out the operator product expansion to the vacuum
condensates adding up to dimension-10 and
 take the assumption of vacuum saturation for the  high
dimension vacuum condensates, they  are always
 factorized to lower condensates with vacuum saturation in the QCD sum rules,
  factorization works well in  large $N_c$ limit.
 In calculation, we observe that the
contributions  from the gluon condensate are suppressed by large
denominators and would not play any significant roles
\cite{Wang1,Wang2,Wang3,Wang4,Wang5}. Furthermore,  we neglect the
terms proportional to the $m_u$ and $m_d$ as their contributions are
of minor importance.

Once analytical results are obtained,   then we can take the
quark-hadron duality and perform Borel transform  with respect to
the variable $P^2=-p^2$, finally we obtain  the following sum rule:
\begin{eqnarray}
\lambda_{Z}^2 e^{-\frac{M_Z^2}{M^2}}= \int_{4m_c^2}^{s_0} ds
\rho(s)e^{-\frac{s}{M^2}} \, ,
\end{eqnarray}
\begin{eqnarray}
\rho(s)&=&\frac{1}{512 \pi^6}
\int_{\alpha_{min}}^{\alpha_{max}}d\alpha
\int_{\beta_{min}}^{1-\alpha} d\beta
\alpha\beta(1-\alpha-\beta)^3(s-\widetilde{m}^2_c)^2(7s^2-6s\widetilde{m}^2_c+\widetilde{m}^4_c)
\nonumber \\
&&+t\frac{ m_c\langle \bar{q}q\rangle}{16 \pi^4}
\int_{\alpha_{min}}^{\alpha_{max}}d\alpha
\int_{\beta_{min}}^{1-\alpha} d\beta
(1-\alpha-\beta)(\alpha+\beta)(s-\widetilde{m}^2_c) (\widetilde{m}^2_c-2s) \nonumber\\
&& +t\frac{ m_c\langle \bar{q}g_s\sigma Gq\rangle}{64 \pi^4}
\int_{\alpha_{min}}^{\alpha_{max}}d\alpha
\int_{\beta_{min}}^{1-\alpha} d\beta (\alpha+\beta)
(3s-2\widetilde{m}^2_c)  \nonumber\\
&&+\frac{m_c^2\langle \bar{q}q\rangle^2}{12 \pi^2}
\int_{\alpha_{mix}}^{\alpha_{max}} d\alpha -\frac{m_c^2\langle
\bar{q}q\rangle \langle \bar{q}g_s \sigma Gq\rangle }{24 \pi^2}
\int_{\alpha_{mix}}^{\alpha_{max}} d\alpha
\left[1+\frac{\widetilde{\widetilde{m}}^2_c}{M^2}\right]\delta
\left(s-\widetilde{\widetilde{m}}^2_c\right) \nonumber\\
&&+\frac{m_c^2\langle \bar{q}g_s \sigma Gq\rangle^2}{192\pi^2
M^6}\int_{\alpha_{mix}}^{\alpha_{max}} d\alpha s^2 \delta
\left(s-\widetilde{\widetilde{m}}^2_c\right) \nonumber\\
&&+\frac{1}{512 \pi^4} \langle \frac{\alpha_s
GG}{\pi}\rangle\int_{\alpha_{min}}^{\alpha_{max}}d\alpha
\int_{\beta_{min}}^{1-\alpha} d\beta
(\alpha+\beta)(1-\alpha-\beta)^2
(10s^2-12s\widetilde{m}^2_c+3\widetilde{m}^4_c) \nonumber\\
&&-\frac{m_c^2}{384 \pi^4} \langle \frac{\alpha_s
GG}{\pi}\rangle\int_{\alpha_{min}}^{\alpha_{max}}d\alpha
\int_{\beta_{min}}^{1-\alpha} d\beta
\frac{(\alpha^3+\beta^3)(1-\alpha-\beta)^3}{\alpha^2 \beta^2} \nonumber\\
&&\left[2s-\widetilde{m}^2_c+ \frac{s^2}{6}\delta(s-\widetilde{m}^2_c)\right] \nonumber\\
&&+\frac{tm_c\langle \bar{q}g_s \sigma Gq\rangle }{384\pi^2}\langle
\frac{\alpha_s GG}{\pi}\rangle
\int_{\alpha_{min}}^{\alpha_{max}}d\alpha
 \left[1+
\frac{\widetilde{\widetilde{m}}^2_c}{M^2}\right] \delta\left(s-\widetilde{\widetilde{m}}^2_c\right)\nonumber\\
&&+\frac{tm_c^3\langle \bar{q}q\rangle }{288\pi^2}\langle
\frac{\alpha_s GG}{\pi}\rangle
\int_{\alpha_{min}}^{\alpha_{max}}d\alpha
\int_{\beta_{min}}^{1-\alpha} d\beta
\frac{(\alpha+\beta)(\alpha^3+\beta^3)(1-\alpha-\beta)}{\alpha^3
\beta^3}\nonumber\\
&& \left[1+
\frac{\widetilde{m}^2_c}{M^2}\right] \delta\left(s-\widetilde{m}^2_c\right)\nonumber\\
&&-\frac{tm_c\langle \bar{q}q\rangle }{96\pi^2}\langle
\frac{\alpha_s GG}{\pi}\rangle
\int_{\alpha_{min}}^{\alpha_{max}}d\alpha
\int_{\beta_{min}}^{1-\alpha} d\beta
\left[1+\frac{(\alpha^3+\beta^3)(1-\alpha-\beta)}{\alpha^2 \beta^2}\right] \left[2+s\delta\left(s-\widetilde{m}^2_c\right)\right] \nonumber\\
&&-\frac{tm_c^3\langle \bar{q}g_s \sigma Gq\rangle }{1152\pi^2
M^4}\langle \frac{\alpha_s GG}{\pi}\rangle
\int_{\alpha_{min}}^{\alpha_{max}}d\alpha
\int_{\beta_{min}}^{1-\alpha} d\beta
\frac{(\alpha+\beta)(\alpha^3+\beta^3)}{\alpha^3 \beta^3} s\delta\left( s-\widetilde{m}^2_c\right) \nonumber\\
&&+\frac{tm_c\langle \bar{q}g_s \sigma Gq\rangle }{384\pi^2}\langle
\frac{\alpha_s GG}{\pi}\rangle
\int_{\alpha_{min}}^{\alpha_{max}}d\alpha
\int_{\beta_{min}}^{1-\alpha} d\beta
\frac{(\alpha^3+\beta^3)}{\alpha^2 \beta^2}\left[1+
\frac{\widetilde{m}^2_c}{M^2}\right] \delta\left(s-\widetilde{m}^2_c\right)\nonumber\\
&&+\frac{m_c^2\langle \bar{q}q\rangle^2 }{72 M^2}\langle
\frac{\alpha_s GG}{\pi}\rangle
\int_{\alpha_{min}}^{\alpha_{max}}d\alpha \left[\frac{1}{\alpha^2}+
\frac{1}{(1-\alpha)^2}\right] \delta\left(s-\widetilde{\widetilde{m}}^2_c\right)\nonumber\\
&&-\frac{m_c^4\langle \bar{q}q\rangle^2 }{216 M^4}\langle
\frac{\alpha_s GG}{\pi}\rangle
\int_{\alpha_{min}}^{\alpha_{max}}d\alpha \left[\frac{1}{\alpha^3}+
\frac{1}{(1-\alpha)^3}\right]
\delta\left(s-\widetilde{\widetilde{m}}^2_c\right)\, ,
\end{eqnarray}
where $\alpha_{max}=\frac{1+\sqrt{1-\frac{4m_c^2}{s}}}{2}$,
$\alpha_{min}=\frac{1-\sqrt{1-\frac{4m_c^2}{s}}}{2}$,
$\beta_{min}=\frac{\alpha m_c^2}{\alpha s -m_c^2}$,
$\widetilde{m}_c^2=\frac{(\alpha+\beta)m_c^2}{\alpha\beta}$,
$\widetilde{\widetilde{m}}_c^2=\frac{m_c^2}{\alpha(1-\alpha)}$,
$t=cos2\theta\in [-1,1]$.

 Differentiating  the Eq.(10) with respect to  $\frac{1}{M^2}$, then eliminate the
 pole residue $\lambda_{Z}$, we can obtain a sum rule for
 the mass of the $Z$,
 \begin{eqnarray}
 M_Z^2= \frac{\int_{4m_c^2}^{s_0} ds
\frac{d}{d \left(-1/M^2\right)}\rho(s)e^{-\frac{s}{M^2}}
}{\int_{4m_c^2}^{s_0} ds \rho(s)e^{-\frac{s}{M^2}}}\, .
\end{eqnarray}

\section{Numerical results and discussions}
The input parameters are taken to be the standard values $\langle
\bar{q}q \rangle=-(0.24\pm 0.01\, \rm{GeV})^3$,  $\langle
\bar{q}g_s\sigma G q \rangle=m_0^2\langle \bar{q}q \rangle$,
$m_0^2=(0.8 \pm 0.2)\,\rm{GeV}^2$, $\langle \frac{\alpha_s
GG}{\pi}\rangle=(0.33\,\rm{GeV})^4 $,
 and $m_c=(1.35\pm0.10)\,\rm{GeV}$ at the energy scale about $\mu=1\, \rm{GeV}$
\cite{SVZ79,Reinders85,Ioffe2005}.

The Belle collaboration  observed the resonance-like structures
$Z(4050)$ and $Z(4250)$ in the $\pi^+\chi_{c1}$ invariant mass
distribution near $4.1 \,\rm{GeV}$ in the exclusive $\bar{B}^0\to
K^- \pi^+ \chi_{c1}$ decays \cite{Belle-chipi}.
 If they are scalar tetraquark states, the central value of the threshold parameter can be taken as
 $s_0=(4.248+0.5)^2\, \rm{GeV}^2\approx 23 \, \rm{GeV}^2$, where we tentatively choose  the
energy gap between the ground states and the first radial excited
states to be $0.5\,\rm{GeV}$.

The present experimental knowledge about the phenomenological
hadronic spectral densities of the tetraquark states is  rather
vague, whether or not there exist   tetraquark states is not
confirmed with confidence, and no knowledge about  the high
resonances; we can borrow some ideas from the baryon spectra
\cite{PDG}.

 For the octet baryons with the quantum numbers $I(J^{P})=\frac{1}{2}({\frac{1}{2}}^+)$,
the mass of the proton (the ground state)  is $M_p=938\,\rm{MeV}$,
and the mass of the first radial excited state $N(1440)$ (the Roper
resonance) is $M_{1440}=(1420-1470)\,\rm{MeV}\approx 1440\,\rm{MeV}$
\cite{PDG}. For the decuplet  baryons with the quantum numbers
$I(J^{P})=\frac{3}{2}({\frac{3}{2}}^+)$ , the mass of  the
$\Delta(1232)$ (the ground state) is
$M_{1232}=(1231-1233)\,\rm{MeV}\approx 1232\,\rm{MeV}$,  and the
mass of the first radial excited state $\Delta(1600)$ is
$M_{1600}=(1550-1700)\,\rm{MeV}\approx 1600\,\rm{MeV}$ \cite{PDG}.
The energy gap  between the ground states and the first radial
excited states can be tentatively taken as $0.5\, \rm{GeV}$ for the
light flavor baryons.

In Ref.\cite{Maiani2008}, Maiani et al assume the $Z(4430)$ is  the
$2S$ $1^{+-}$  hidden charm tetraquark state to take into account
the decay mode $Z(4430)\rightarrow \psi(2S)+\pi^+$, as  the $1S$
$1^{+-}$  hidden charm tetraquark states lie at the interval
$(3750-3880)\,\rm{MeV}$ and have the decay mode $X^+(1S)\rightarrow
\psi(1S)+\pi^+$ or $\eta_c(1S)+\rho^+$ in the constituent diquark
model \cite{Maiani20042}. The energy gap  between the ground state
and the first radial excited state is estimated to be
$M_{\psi(2S)}-M_{\psi(1S)}\sim 0.59 \,\rm{GeV}$ for the heavy
tetraquark states.

We take it for granted that the energy gap  between the ground
states and the first radial excited states is about $0.5\,\rm{GeV}$,
and use this value as a guide to determine the threshold parameter
$s_0$ with the QCD sum rules.

We explore
 whether or not there exist  scalar tetraquark states
which consist of a scalar (pseudoscalar) diquark-antidiquark pair at
the energy interval $(4.1-4.3)\,\rm{GeV}$, and choose the larger
value $s_0=(4.248+0.5)^2\, \rm{GeV}^2\approx 23 \, \rm{GeV}^2$
rather than the smaller value $s_0=(4.051+0.5)^2\, \rm{GeV}^2\approx
22 \, \rm{GeV}^2$ to take into account all possible contributions
from the ground states.

 In the conventional QCD sum
rules \cite{SVZ79,Reinders85}, there are two criteria (pole
dominance and convergence of the operator product expansion) for
choosing  the Borel parameter $M^2$ and threshold parameter $s_0$
\footnote{For the tetraquark states consist of light flavors, if the
perturbative terms have the main contribution (in the conventional
QCD sum rules, the perturbative terms always have the main
contribution), we can approximate the spectral density with the
perturbative term (where the $A$ are some numerical coefficients)
\cite{Wang0708},
\begin{eqnarray}
B_M\Pi \sim A \int_0^\infty s^4 e^{-\frac{s}{M^2}}ds=A
M^{10}\int_0^\infty t^4 e^{-t}dt \, ,
\end{eqnarray}
then take the pole dominance condition,
\begin{eqnarray}
\frac{\int_0^{t_0} t^4 e^{-t}dt}{\int_0^\infty t^4 e^{-t}dt}\geq
50\% \, ,
\end{eqnarray}
and obtain the approximated  relation,
\begin{eqnarray}
t_0&=&\frac{s_0}{M^2}\geq 4.7 \, .
\end{eqnarray}

The superpositions of different interpolating currents can only
change the contributions from different terms in the operator
product expansion, and  improve convergence, they cannot change the
leading behavior of the spectral density $\rho(s)\propto s^4 $ of
the perturbative term \cite{Wang0708}.

This relation is difficult to satisfy for the light flavor
tetraquark states \cite{Wang1,Wang2,Wang3,Wang4,Wang5}, if we take
the Borel parameter has the typical value $M^2=1\, \rm{GeV}^2$,
$s_0\geq 4.7\, \rm{GeV}^2$, the threshold parameter is too large for
the light tetraquark state candidates $f_0(980)$, $a_0(980)$, etc.

  The hidden charm (or bottom) tetraquark states and open
bottom tetraquark states may satisfy the relation, as they always
have larger Borel parameter $M^2$ and threshold parameter $s_0$
\cite{Narison07,Lee07-sum,SR4430,SR4260}. Their  spectral densities
have the form $\rho(s)=C_1 s^4+C_2s^3+C_3s^2+\cdots$, where the
$C_i$ are coefficients, and exhibit  the same leading behavior
$\rho(s)\propto s^4 $ as the light flavor tetraquark sates. If we
take $M^2=1\,\rm{GeV}^2$, $s_0\geq 4.7\, \rm{GeV}^2$, the threshold
parameter $s_0$ is  too low for the hidden charm or open bottom
tetraquark states, there is  a large  room for choosing  larger
threshold parameter to take into account the ground state
contribution. We draw the conclusion that the hidden charm (bottom)
tetraquark states and open bottom tetraquark states have possibility
to satisfy the pole dominance condition.

 In this article, the vacuum condensate of
the highest dimension $\langle \bar{q} g_s \sigma G q\rangle^2$
serve as a criterion for choosing the Borel parameter $M^2$. At the
value $M^2_{min} \geq 2.2\,\rm{GeV}^2$, its contribution is less
than $10\%$ (see Fig.3), we expect the operator product expansion is
convergent. The relation in Eq.(15) indicates $s_0 \geq
10.5\,\rm{GeV}^2$, if we take a large Borel parameter $M^2\geq
2M_{min}^2$, then $s_0 \geq 21\,\rm{GeV}^2$, our phenomenological
estimation  $s_0\sim 23\,\rm{GeV}^2$ is  reasonable.}.

\begin{figure}
 \centering
 \includegraphics[totalheight=6cm,width=7cm]{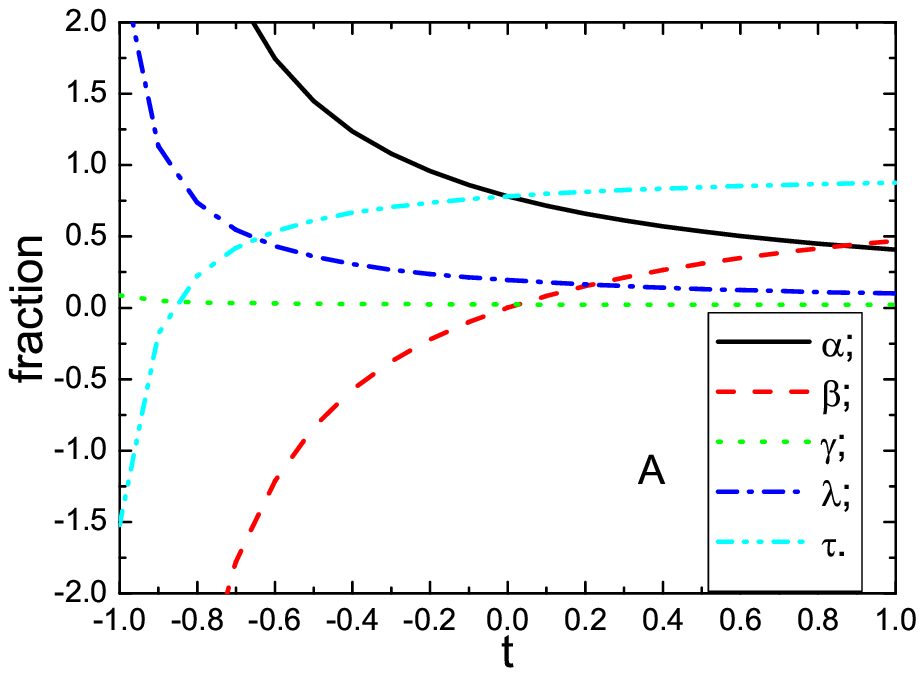}
 \includegraphics[totalheight=6cm,width=7cm]{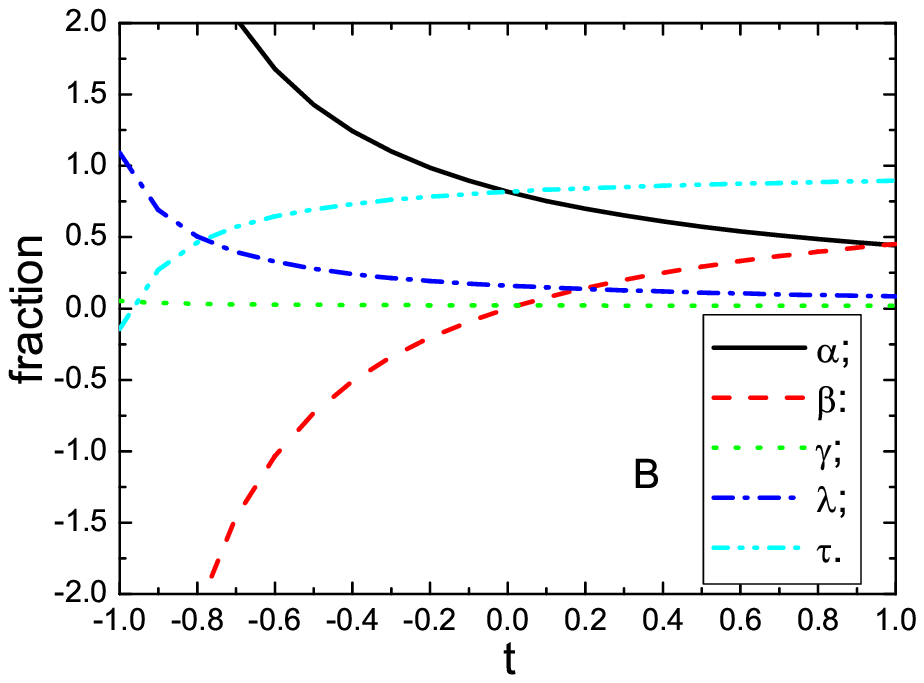}
   \caption{ The contributions from different terms with variation of the  parameter $t$. The $A$ and  $B$ correspond to the
   threshold parameters $s_0=23 \,\rm{GeV}^2$ and $25\,\rm{GeV}^2$ respectively. The notations
   $\alpha$, $\beta$, $\gamma$, $\lambda$ and $\tau$  correspond to
    the perturbative term, $\langle
\bar{q} q \rangle +\langle \bar{q}g_s\sigma G q \rangle$ term,
$\langle \frac{\alpha_s GG}{\pi} \rangle $+$\langle \frac{\alpha_s
GG}{\pi} \rangle \left[\langle \bar{q} q \rangle +\langle
\bar{q}g_s\sigma G q \rangle+ \langle \bar{q} q \rangle^2\right]$
term, $\langle \bar{q} q \rangle^2$ + $\langle \bar{q} q
\rangle\langle \bar{q}g_s\sigma G q \rangle$+ $\langle
\bar{q}g_s\sigma G q \rangle^2$ term and perturbative +$\langle
\bar{q} q \rangle +\langle \bar{q}g_s\sigma G q \rangle$ term,
respectively. Here we take $M^2=3\,\rm{GeV}^2$ and the central
values of other input parameters.}
\end{figure}

\begin{figure}
 \centering
 \includegraphics[totalheight=7cm,width=8cm]{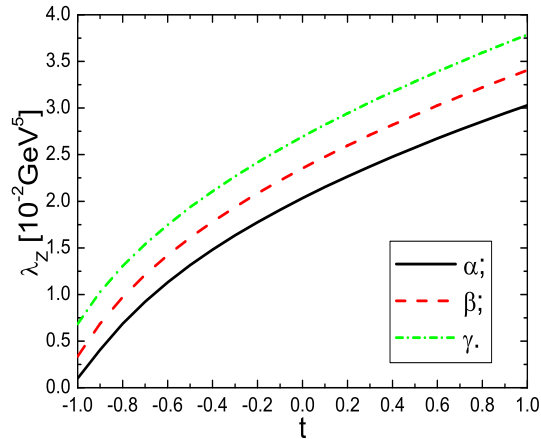}
   \caption{ The pole residue (or coupling) with variation of the  parameter $t$. The notation $\alpha$, $\beta$ and
   $\gamma$ correspond to the threshold parameters $s_0=23\,\rm{GeV}^2$, $24\,\rm{GeV}^2$ and $25\,\rm{GeV}^2$, respectively.
   Here
we take $M^2=3\,\rm{GeV}^2$ and the central values of other input
parameters.. }
\end{figure}

The contributions from different terms  with variation of the
parameter $t$ in the operator product expansion are shown in Fig.1.
From the figure, we can see that the contributions from the term
$\langle \bar{q} q \rangle +\langle \bar{q}g_s\sigma G q \rangle$
are negative at the interval $-1\leq t <0$, which cancel out the
contribution from the perturbative term greatly. The net
contributions from the perturbative term+$\langle \bar{q} q \rangle
+\langle \bar{q}g_s\sigma G q \rangle$ increase with  variation of
the $t$, and reach the largest value at  $t=1$. The contributions
from the gluon condensates $\langle \frac{\alpha_s GG}{\pi} \rangle
$+$\langle \frac{\alpha_s GG}{\pi} \rangle \left[\langle \bar{q} q
\rangle +\langle \bar{q}g_s\sigma G q \rangle+ \langle \bar{q} q
\rangle^2\right]$ are very small and decrease with the parameter $t$
monotonously;  the contributions from the  high dimension
condensates $\langle \bar{q} q \rangle^2$ + $\langle \bar{q} q
\rangle\langle \bar{q}g_s\sigma G q \rangle$+ $\langle
\bar{q}g_s\sigma G q \rangle^2$  also decrease with the parameter
$t$ monotonously. In other words, the operator product expansion
converges more quickly for larger $t$ at the interval $t\in [-1,1]$,
   we can choose the value $t=1$.

   On the other hand, the coupling of the
interpolating current $J(x)$ to the tetraquark state becomes
stronger with larger $t$, see Fig.2. It is reasonable to take the
interpolating current with the strongest coupling to the tetraquark
state.

In Figs.3-4, we plot the contributions from different terms in the
operator product expansion. The contribution from the term $\langle
\frac{\alpha_s GG}{\pi}\rangle$ is tiny and can be safely neglected.
The contributions from the terms involving the gluon condensates are
less than $8\%$ even at very small Borel parameter $M^2$, the gluon
condensate plays a minor important role. The vacuum condensate of
the highest dimension $\langle \bar{q} g_s \sigma G q\rangle^2$
serve as a criterion for choosing the Borel parameter $M^2$. At the
value $M^2_{min} \geq 2.2\,\rm{GeV}^2$, its contribution is less
than $10\%$, we expect the operator product expansion is convergent.

The contributions from the vacuum condensates of high dimension
$\langle \bar{q}q\rangle^2+\langle \bar{q}q\rangle\langle \bar{q}g_s
\sigma Gq\rangle$ vary with the threshold parameter $s_0$ remarkably
and serve as a criterion for choosing the threshold parameter $s_0$.
At the value $s_0\geq 23\,\rm{GeV}^2$, their contributions are less
than (or equal) $ 10 \%$ (see Fig.3-A), we expect the operator
product expansion is convergent. The contributions from the vacuum
condensates $\langle \bar{q}q\rangle^2+\langle
\bar{q}q\rangle\langle \bar{q}g_s \sigma Gq\rangle+\langle
\bar{q}g_s \sigma Gq\rangle^2$ are less than  $13.5 \%$ at the
values $M^2\geq 2.2 \, \rm{GeV}^2$ and $s_0 \geq 23\,\rm{GeV}^2$.
The contributions from the vacuum condensates $\langle \bar{q} q
\rangle^2$ + $\langle \bar{q} q \rangle\langle \bar{q}g_s\sigma G q
\rangle$+ $\langle \bar{q}g_s\sigma G q \rangle^2$+ $\langle
\frac{\alpha_s GG}{\pi} \rangle $+$\langle \frac{\alpha_s GG}{\pi}
\rangle \left[\langle \bar{q} q \rangle +\langle \bar{q}g_s\sigma G
q \rangle+ \langle \bar{q} q \rangle^2\right]$ are less than $18\%$,
the main contributions come from the perturbative term +$\langle
\bar{q} q \rangle$ +$\langle \bar{q}g_s\sigma G q \rangle$, see
Fig.4. The operator product expansion is convergent at the values
$M^2_{min}\geq 2.2 \, \rm{GeV}^2$ and $s_0 \geq 23\,\rm{GeV}^2$.

\begin{figure}
 \centering
 \includegraphics[totalheight=5cm,width=6cm]{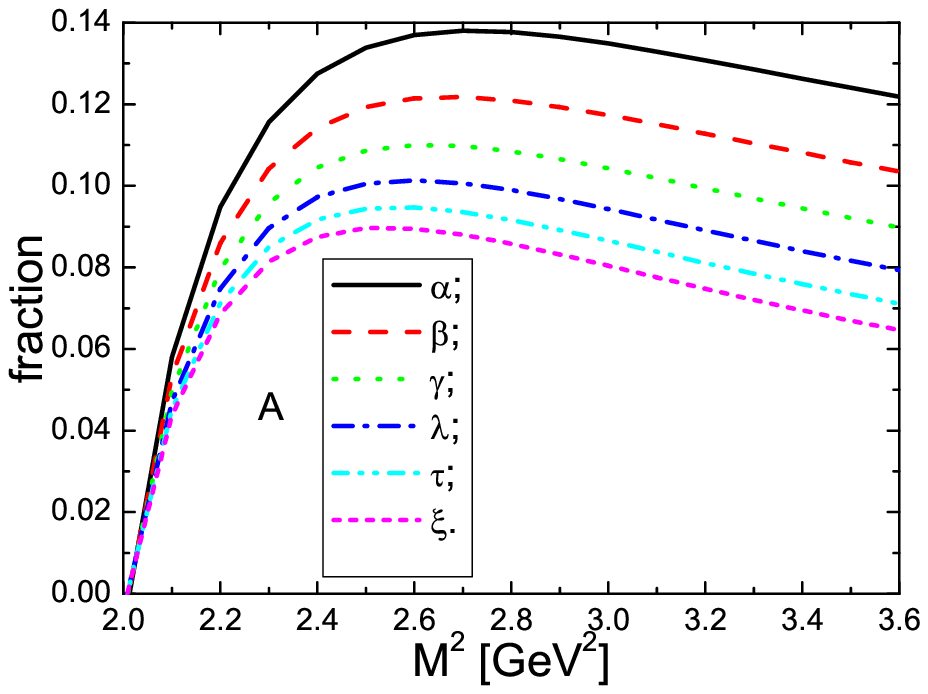}
 \includegraphics[totalheight=5cm,width=6cm]{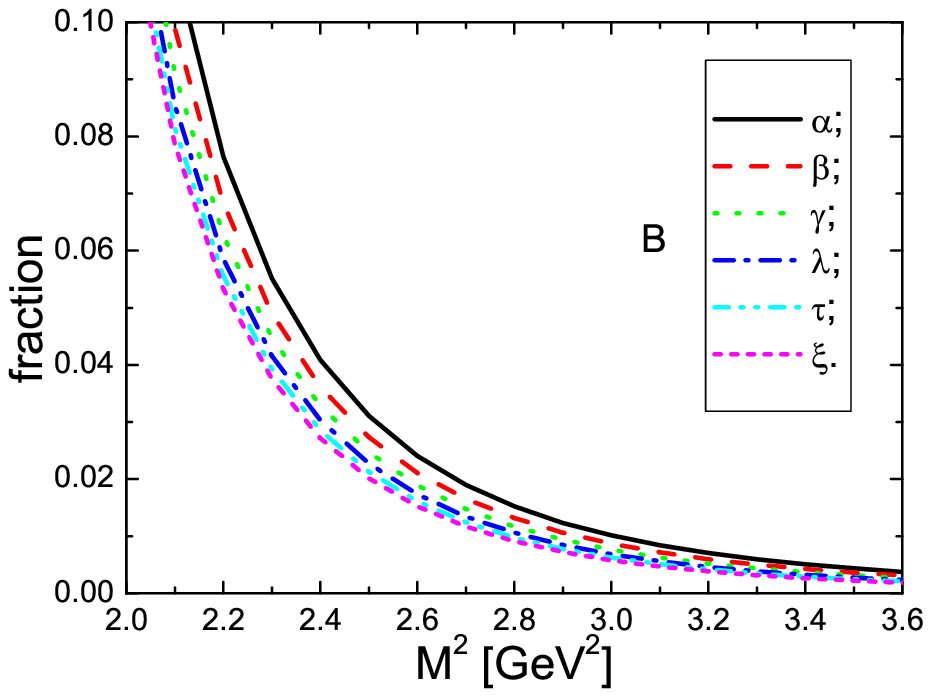}
 \includegraphics[totalheight=5cm,width=6cm]{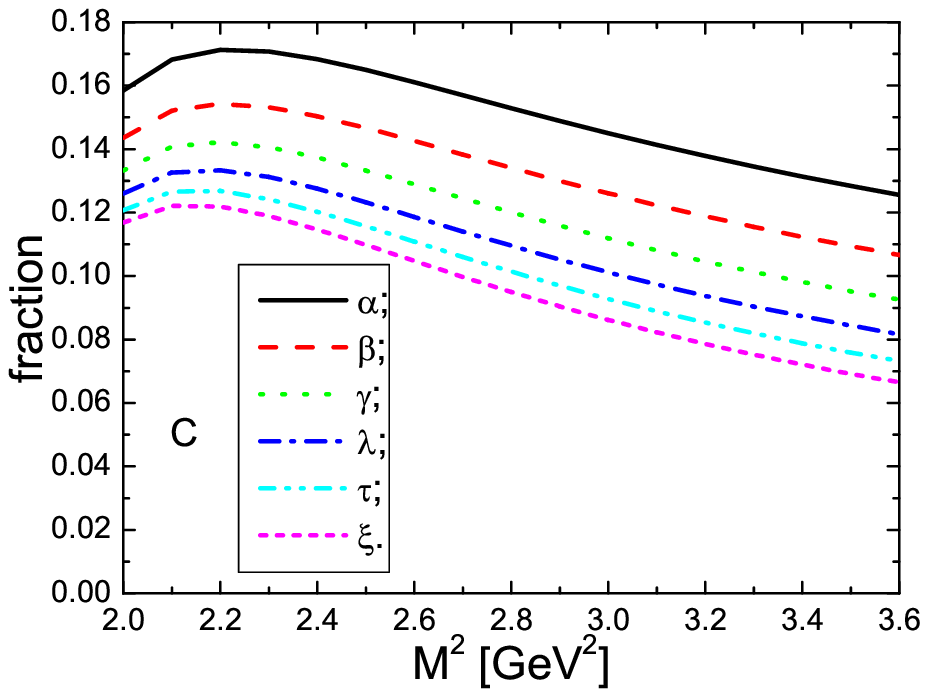}
 \includegraphics[totalheight=5cm,width=6cm]{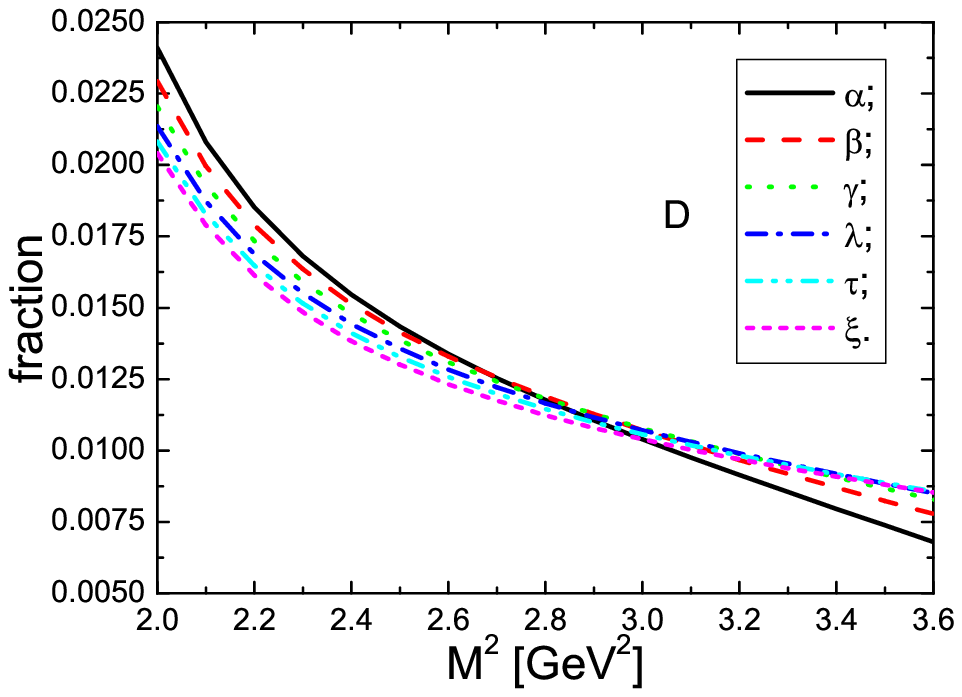}
 \includegraphics[totalheight=5cm,width=6cm]{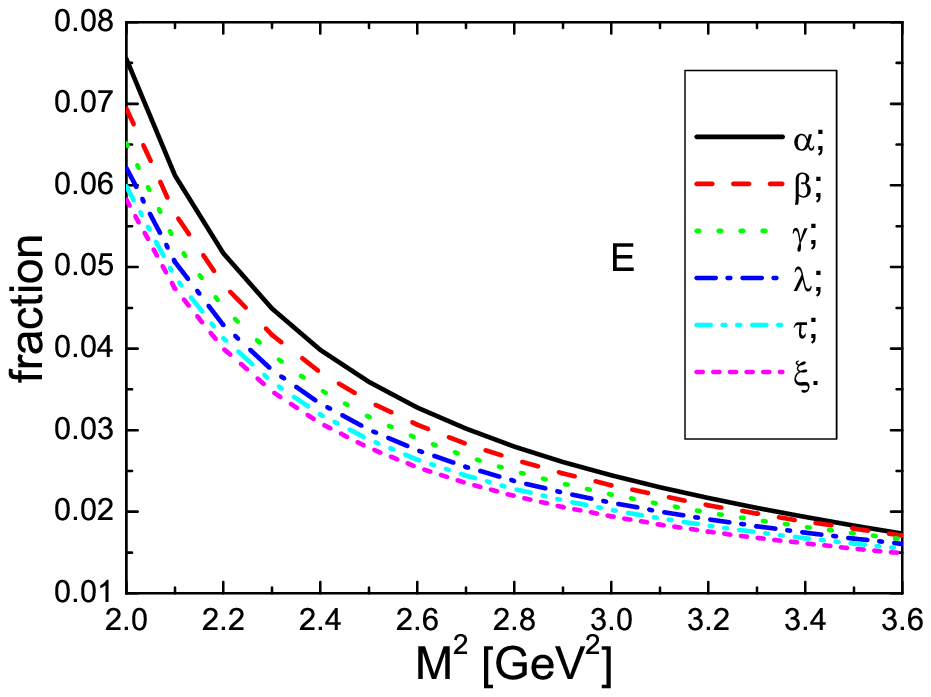}
 \includegraphics[totalheight=5cm,width=6cm]{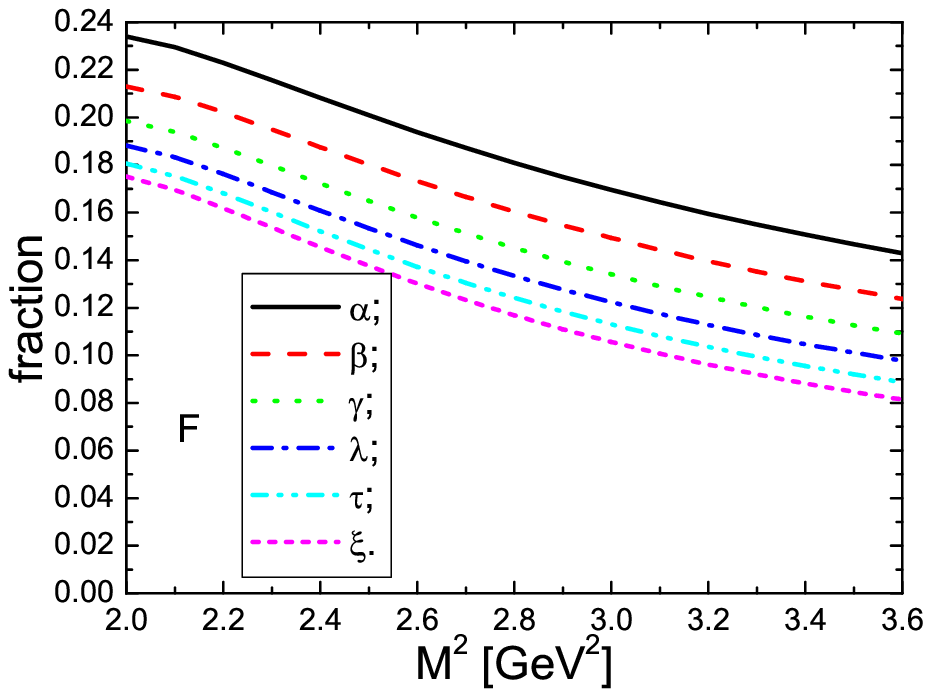}
   \caption{ The contributions from the vacuum condensates  with variation of the Borel parameter $M^2$. The $A$,
   $B$, $C$, $D$, $E$ and $F$ correspond to the contributions from
   the $\langle \bar{q} q \rangle^2$ +$\langle \bar{q} q \rangle\langle \bar{q}g_s\sigma G q \rangle$ term,
$\langle \bar{q}g_s\sigma G q \rangle^2$ term, $\langle \bar{q} q
\rangle^2$ +$\langle \bar{q} q \rangle\langle \bar{q}g_s\sigma G q
\rangle$ + $\langle \bar{q}g_s\sigma G q \rangle^2$ term, $\langle
\frac{\alpha_s GG}{\pi} \rangle $ term, $\langle \frac{\alpha_s
GG}{\pi} \rangle $+$\langle \frac{\alpha_s GG}{\pi} \rangle
\left[\langle \bar{q} q \rangle +\langle \bar{q}g_s\sigma G q
\rangle+ \langle \bar{q} q \rangle^2\right]$ term and $\langle
\bar{q} q \rangle^2$ + $\langle \bar{q} q \rangle\langle
\bar{q}g_s\sigma G q \rangle$+ $\langle \bar{q}g_s\sigma G q
\rangle^2$+ $\langle \frac{\alpha_s GG}{\pi} \rangle $+$\langle
\frac{\alpha_s GG}{\pi} \rangle \left[\langle \bar{q} q \rangle
+\langle \bar{q}g_s\sigma G q \rangle+ \langle \bar{q} q
\rangle^2\right]$ term, respectively.     The notations
   $\alpha$, $\beta$, $\gamma$, $\lambda$, $\tau$ and $\xi$ correspond to the threshold
   parameters $s_0=20\,\rm{GeV}^2$,
   $21\,\rm{GeV}^2$, $22\,\rm{GeV}^2$, $23\,\rm{GeV}^2$, $24\,\rm{GeV}^2$ and $25\,\rm{GeV}^2$, respectively.
    Here we take  $t=1$
and the central values of other input parameters. }
\end{figure}

\begin{figure}
 \centering
 \includegraphics[totalheight=6cm,width=7cm]{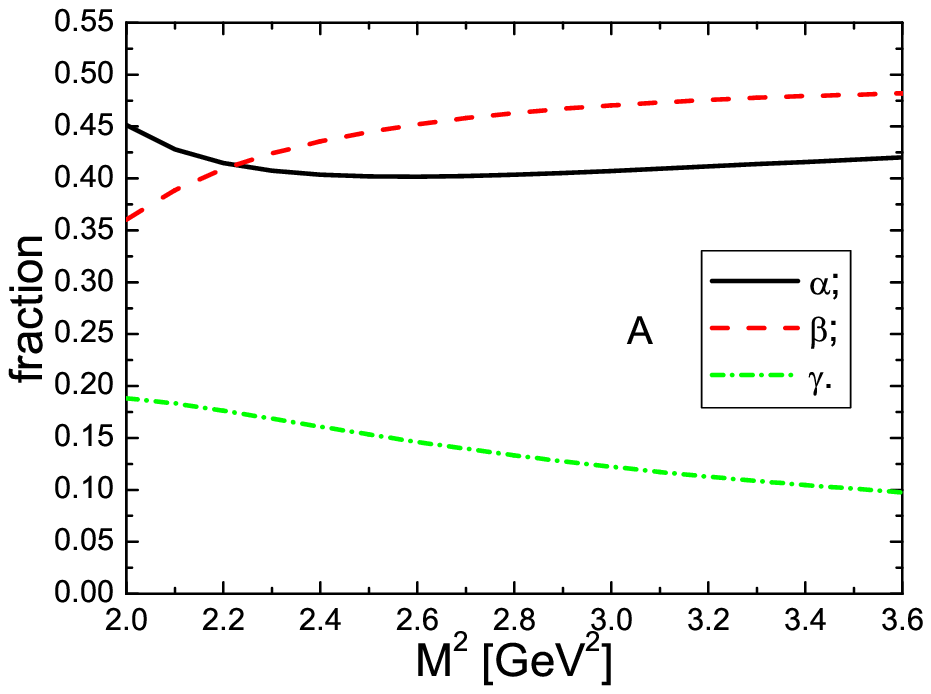}
 \includegraphics[totalheight=6cm,width=7cm]{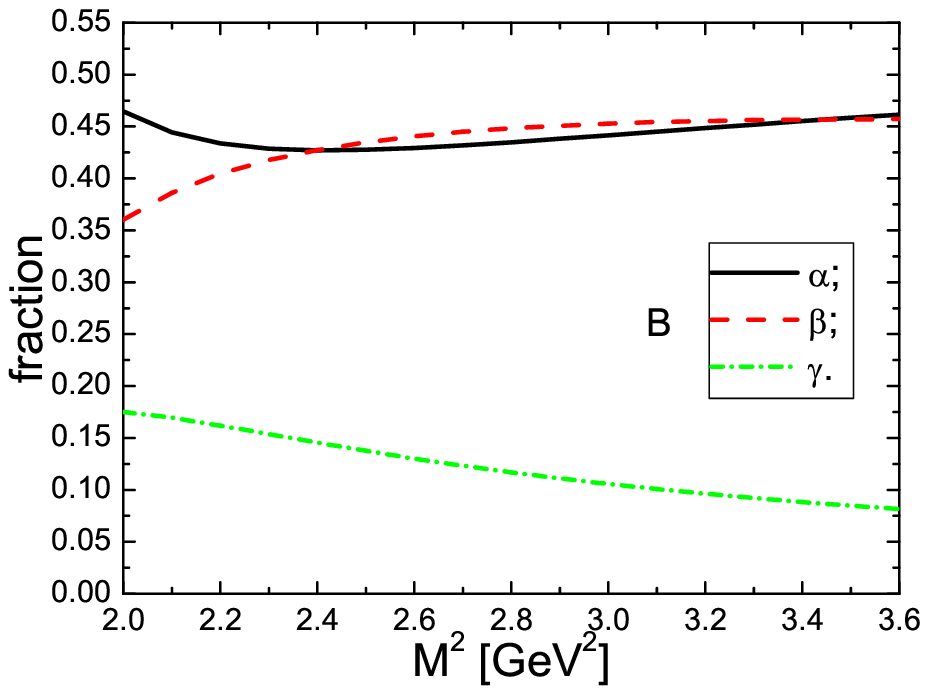}
   \caption{ The contributions from different terms with variation of the Borel parameter $M^2$. The $A$ and $B$
correspond    to  the threshold parameters $s_0=23\,\rm{GeV}^2$ and
   $25\,\rm{GeV}^2$ respectively.
   The notations
   $\alpha$, $\beta$ and $\gamma$ correspond to  to the perturbative term, $\langle
\bar{q} q \rangle$ +$\langle \bar{q}g_s\sigma G q \rangle$ term and
$\langle \bar{q} q \rangle^2$ + $\langle \bar{q} q \rangle\langle
\bar{q}g_s\sigma G q \rangle$+ $\langle \bar{q}g_s\sigma G q
\rangle^2$+ $\langle \frac{\alpha_s GG}{\pi} \rangle $+$\langle
\frac{\alpha_s GG}{\pi} \rangle \left[\langle \bar{q} q \rangle
+\langle \bar{q}g_s\sigma G q \rangle+ \langle \bar{q} q
\rangle^2\right]$ term, respectively. Here we take
 $t=1$ and the central values of other input
parameters.}
\end{figure}

In Fig.5, we plot the contribution from the pole term with variation
of the threshold parameter $s_0$. For the central values of the
input parameters (except for $t=1$), the contribution from the pole
term is larger than $50\%$ at the values $M^2_{max} \leq 3.2 \,
\rm{GeV}^2 $ and $s_0\geq 23\, \rm{GeV}^2$.

In this article, the threshold   parameter and the Borel parameter
are taken as $s_0=(24 \pm 1)\,\rm{GeV}^2$ and
$M^2=(2.2-3.2)\,\rm{GeV}^2$ respectively, the  contribution from the
pole term is about $(51-88)\%$, the two criteria of the QCD sum
rules are full filled \cite{SVZ79,Reinders85}. We can take smaller
Borel parameter and threshold parameter to satisfy the two criteria
of the QCD sum rules marginally, however, the Borel window is rather
small, $M^2_{max}-M^2_{min}< 1\,\rm{GeV}^2$.

Taking into account all uncertainties of the input parameters,
finally we obtain the values of the mass and pole reside of
 the   $Z$, which are  shown in Figs.4-5. From the figures, we can
 see that at the value $M^2\leq 2.6\,\rm{GeV}^2$,  the mass and the pole
 residue change remarkably with variation of the Borel parameter, we
 take the value $M^2=(2.6-3.2)\,\rm{GeV}^2$, and obtain
\begin{eqnarray}
M_{Z}&=&(4.36\pm0.18)\,\rm{GeV} \, ,  \nonumber\\
\lambda_{Z}&=&(3.38\pm0.65)\times 10^{-2}\,\rm{GeV}^5 \,   .
\end{eqnarray}
The  meson  $Z(4250)$   may be a scalar tetraquark state, other
possibilities, such as a hadro-charmonium resonances and a
$D_1^+\bar{D}^0+ D^+\bar{D}_1^0$ molecular states  are not excluded.

\begin{figure}
 \centering
 \includegraphics[totalheight=7cm,width=8cm]{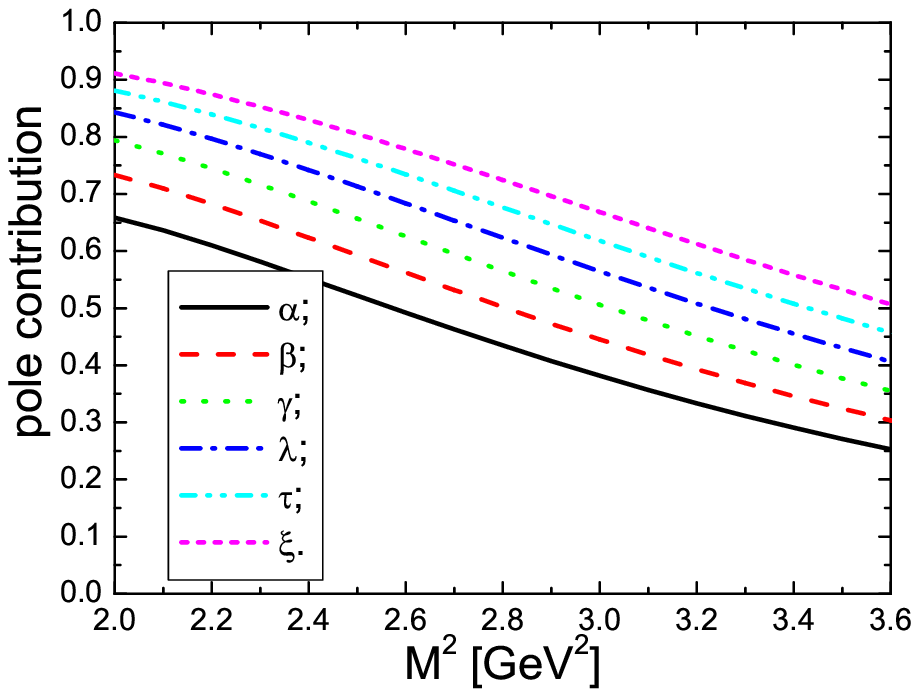}
   \caption{ The contribution from the pole term  with variation of the  Borel parameter $M^2$. The notations
   $\alpha$, $\beta$, $\gamma$, $\lambda$, $\tau$ and $\xi$ correspond to the threshold parameters $s_0=20\,\rm{GeV}^2$,
   $21\,\rm{GeV}^2$, $22\,\rm{GeV}^2$, $23\,\rm{GeV}^2$, $24\,\rm{GeV}^2$ and $25\,\rm{GeV}^2$, respectively.   }
\end{figure}

\begin{figure}
 \centering
 \includegraphics[totalheight=7cm,width=8cm]{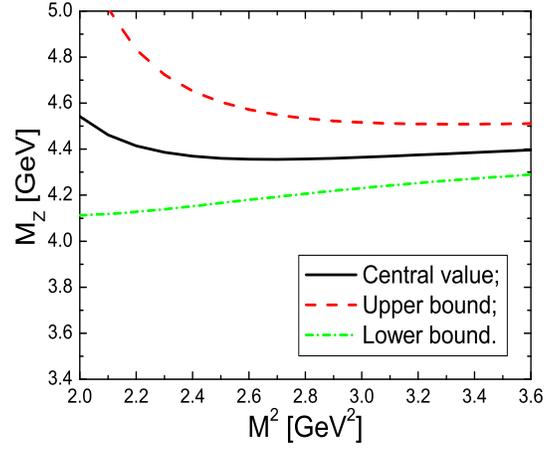}
   \caption{ The mass $M_{Z}$ with variation of the  Borel parameter $M^2$. }
\end{figure}

\begin{figure}
 \centering
 \includegraphics[totalheight=7cm,width=8cm]{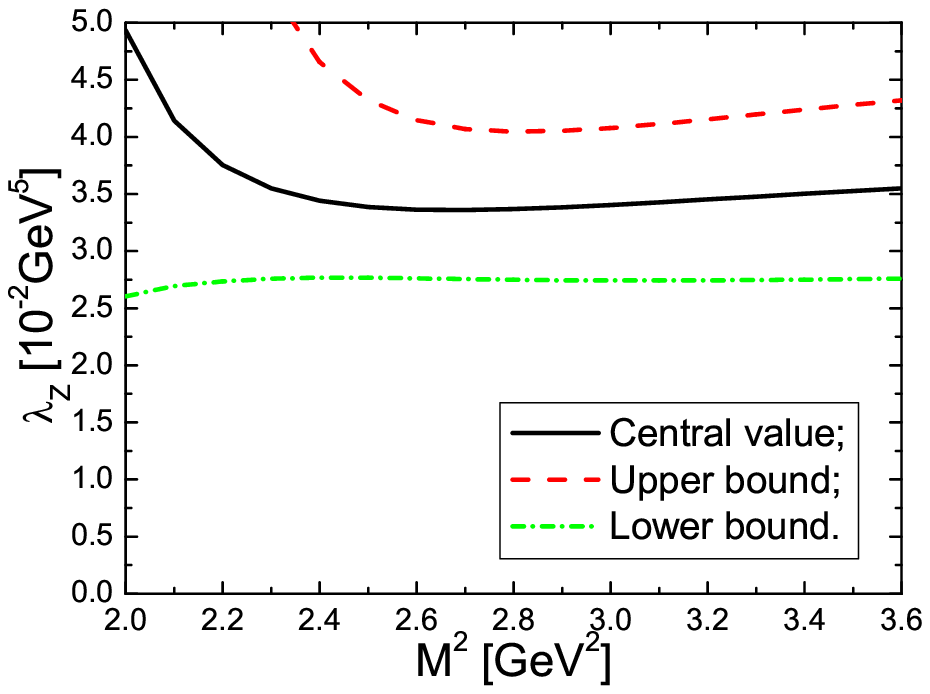}
   \caption{ The pole residue $\lambda_{Z}$ with variation of the  Borel parameter $M^2$. }
\end{figure}

The $Z(4250)$  lie about $(0.5-0.6)\,\rm{GeV}$ above the
$\pi^+\chi_{c1}$ threshold, if it is  a tetraquark state, the decay
$ Z \to \pi^+\chi_{c1}$ can take place with the OZI super-allowed
"fall-apart" mechanism, which can take into account the large total
width naturally; on the other hand, if it is a $D_1^+\bar{D}^0+
D^+\bar{D}_1^0$ molecular state, the decay can occur through the
final-state re-scattering effects, $ Z \to D_1^+\bar{D}^0+
D^+\bar{D}_1^0\to \pi^+\chi_{c1}$, and the corresponding  width may
be narrow, we have to search for other decay channels to accommodate
the large total width.

The typical decay mode $Z \to D^+ \bar{D}^0$ is  kinematically
allowed,  we can determine the spins of the $Z(4250)$ with the
angular distribution of the final state $D^+\bar{D}^0$. If the decay
$Z \to D^+ \bar{D}^0$ is not observed (or the width is rather
narrow), the $Z(4250)$ may be a hadro-charmonium resonance (bound
state of a relatively compact charmonium ($\chi_{c1}$) inside a
light hadron ($\pi^+$) having a larger spatial size)
\cite{Voloshin07}. The decay $ Z \to \pi^+\chi_{c1}$ occurs with the
"fall-apart" mechanism and the width  is large; while the decay $Z
\to D^+ \bar{D}^0$ takes place through the final-state re-scattering
effects ($Z \to \pi^+ \chi_{c1} \to D^+ \bar{D}^0$) and the width
may be  narrow.

\section{Conclusion}
In this article, we assume that there exists a scalar hidden charm
tetraquark state in the $\pi^+ \chi_{c1}$ invariant mass
distribution, and study its mass using the QCD sum rules. The
numerical result   indicates that the mass is about
  $M_{Z}=(4.36\pm0.18)\,\rm{GeV}$, which
 is  consistent with the experimental data.
    The hidden charm meson
$Z(4250)$   may be a tetraquark state. Other possibilities, such as
a  hadro-charmonium resonance  and a $D_1^+\bar{D}^0+
D^+\bar{D}_1^0$ molecular state  are not excluded; more experimental
data are still needed to identify it.

\section*{Acknowledgements}
This  work is supported by National Natural Science Foundation,
Grant Number 10775051, and Program for New Century Excellent Talents
in University, Grant Number NCET-07-0282.

\end{document}